\documentclass[aps, preprintnumbers,prd, amsmath,amssymb, twocolumn]{revtex4-1} %notitlepage, ,10pt
\usepackage{graphicx}% Include figure files latexsym,array,enumerate,letter,numbers,
\usepackage{bm}% bold math
\usepackage{color}
\usepackage{epsfig}
\usepackage{mathtools}
\usepackage{cases}
\usepackage{booktabs}
\usepackage{multirow}

\usepackage[
colorlinks=true,
filecolor=black,
anchorcolor=blue,
linkcolor=blue,
citecolor=cyan, %red,
urlcolor=cyan,
linktocpage=true,
plainpages=false,
breaklinks=true,
pdfstartview=FitH
]{hyperref}
\def\di{\mathrm{d}}
\def\P{P}

\def\pa{{\phi}}
\def\pb{\tilde{\phi}}
\def\x{\mathbf{x}}
\def\d{\mathrm{d}}

\def\fy{f_{\text{yr}}}

\def\l{\left(}
\def\r{\right)}

\newcommand{\dd}{\mathrm{d}}
\newcommand{\mpl}{\ensuremath{M_{P}}}
\newcommand{\rh}{\rho}
\newcommand{\si}{\sigma}
\newcommand{\M}{M}
\newcommand{\D}{\nabla}
\newcommand{\bx}{\ensuremath{{\x}}}
\newcommand{\rhoDM}{\rho_{\text{DM}}}
\newcommand{\vep}{\varepsilon}
\newcommand{\vx}{\mathbf{x}}

\begin{document}
\title{Pulsar Timing Residual induced by Wideband Ultralight Dark Matter with Spin 0,1,2}
\author{Sichun Sun$^{1}$} \email[ ]{sichunssun@bit.edu.cn}
%\thanks{(co-first author)}
\author{Xing-Yu Yang$^{2,3}$} \email[ ]{yangxingyu@itp.ac.cn}
%\thanks{(co-first author)}
\author{Yun-Long Zhang$^{4,5,6}$}\email[]{zhangyunlong@nao.cas.cn}
%\thanks{(corresponding author)}
\affiliation{$^1$School of Physics, Beijing Institute of Technology, Haidian District, Beijing 100081, China}
\affiliation{$^2$CAS Key Laboratory of Theoretical Physics, Institute of Theoretical Physics, Chinese Academy of Sciences, Beijing 100190, China}
\affiliation{$^3$School of Physical Sciences, University of Chinese Academy of Sciences, Beijing 100049, China}
\affiliation{$^4$National Astronomy Observatories, Chinese Academy of Science, Beijing, 100101, China}
\affiliation{$^5$School of Fundamental Physics and Mathematical Sciences, Hangzhou Institute for Advanced Study, University of Chinese Academy of Sciences, Hangzhou 310024, China}
\affiliation{$^6$International Center for Theoretical Physics Asia-Pacific, Beijing/Hangzhou, China}

\begin{abstract}
The coherent oscillation of ultralight dark matter in the mass regime around $10^{-23}$ eV  induces changes in gravitational potential with the frequency in the nanohertz range. This effect is known to produce a monochromatic signal in the pulsar timing residuals. Here we discuss a multifield scenario that produces a wide spectrum of frequencies, such that the ultralight particle oscillation can mimic the pulsar timing signal of stochastic common spectrum process. We discuss how ultralight dark matter with various spins produces such a wide band spectrum on pulsar timing residuals and perform the Bayesian analysis to constrain the parameters. It turns out that the stochastic background detected by NANOGrav can be associated with a wideband ultralight dark matter.
\end{abstract}
\date{Augest 13, 2022}
%\date{January 15, 2022}
\maketitle
\tableofcontents

\section{Introduction}
The pulsar timing array (PTA) observations are sensitive to the gravitational waves of frequencies around the nanohertz range, including EPTA \cite{Lentati:2015qwp}, PPTA \cite{Shannon:2015ect} and NANOGrav \cite{Arzoumanian:2018saf} experiments. The PTA observations can probe various astrophysical phenomena including gravitational waves from, e.g., supermassive black hole binaries (SMBHs), inflation, cosmic strings, modified gravity, phase transitions etc \cite{Anholm:2008wy}. Interestingly, the recent NANOGrav experiment has reported some evidence for a stochastic common-spectrum process after the analysis of 12.5 year dataset \cite{Alam:2020fjy, NANOGrav:2020qll, NANOGrav:2020bcs}, which has also been supported in the analysis of PPTA and EPTA datasets  \cite{Goncharov:2021oub, Chen:2021rqp}.

Significantly, such kind of common spectrum can be interpreted as stochastic gravitational waves background (SGWB) from various sources \cite{Ellis:2020ena,Bian:2020urb}. If the signal is from the  stochastic gravitational waves background, then the characteristic strain of gravitational waves $h_c(f)$ has the amplitude of order $10^{-15}$ around the frequency $\fy \equiv \text{yr}^{-1} \simeq 31.7$ nHz, with an extended frequency spectrum \cite{NANOGrav:2020bcs}. The correlator of the timing residuals is $\langle{R_a R_b}\rangle =\Gamma_{ab} \int{\di} f S_{c}(f)$, where $\Gamma_{ab}$ is the overlap reduction function. The spectral density is $S_{c}(f) = \frac{h_c(f)^2}{12\pi^2 f^3}$. The measured timing residual of the common spectrum in frequency domain is
    \begin{align}\label{scpf}
R_{c}(f) &\equiv \sqrt{\frac{S_{c}(f)}{T_\text{s}}}=\frac{1}{\sqrt{3}} \frac{h_c(f)}{2\pi f}\left(\frac{f_{\text{s}}}{ f}\right)^{ 1/2},
    \end{align}
where $T_{\text{s}}$ is the span between the  maximum and minimum times of arrival in the pulsar timing array, and $f_{\text{s}}\equiv 1/T_{\text{s}}$.
%(e.g. $T_{\text{s}} \simeq12.8$yr for the Nanograv 12.5-year dataset, )

PTA is also sensitive to the nongravitational wave signals, such as the ultralight dark matter through the timing residuals \cite{Khmelnitsky:2013lxt,Porayko:2014rfa,Porayko:2018sfa}. Oscillating ultralight bosons across the cosmic space are interesting dark matter candidates \cite{Hu:2000ke,Hui:2016ltb, Arias:2012az}, which can be either axionlike pseudoscalars or scalar fields such as moduli \cite{Preskill:1982cy,Abbott:1982af,Dine:1982ah,Svrcek:2006yi}. The ultralight dark photon field (vector) oscillation \cite{Nelson:2011sf, Salehian:2020asa,  Nomura:2019cvc}, as well as the dark massive graviton oscillation as dark matter from bimetric gravity \cite{Aoki:2016zgp,Babichev:2016hir,Marzola:2017lbt, Armaleo:2019gil,Armaleo:2020yml,Armaleo:2020efr}, produced during inflation were also proposed.

The ultralight dark matter can be described by a collection of plane waves with energy ${m}c^2$ and momentum ${m} v$, with the typical velocity $v\simeq 10^{-3} c$ in the halo. The de Broglie wavelength is
$
\lambda_{\text {dB}}=\frac{2\pi\hbar}{{m} v}
\simeq 4 \text{kpc}
\left(\frac{10^{-23}\text{eV}}{m}\right)  \left(\frac{10^{-3}}{v}\right)$.
Especially, the ultralight dark matter with the mass around $10^{-23}$ eV, also known as the fuzzy dark matter, was suggested to resolve small-scale problems of the cold collisionless dark matter models \cite{Hu:2000ke,Hui:2016ltb}. 
However, if the single scalar field accounts for the whole dark matter density in the Galaxy, then these explanations are in tension with the latest Lyman-alpha forest constraints with the mass below $2 \times 10^{-20}$ eV~\cite{Irsic:2017yje,Armengaud:2017nkf,Kobayashi:2017jcf,Rogers:2020ltq,Hui:2021tkt}. The bound becomes more stringent if accounting for the quantum pressure \cite{Zhang:2017chj, Nori:2018pka}, and can also be affected by considering different self-interactions \cite{Desjacques:2017fmf}. For the wideband case with a spread mass spectrum, there have not been systematic studies on the simulation. The constraints on the vector(dark photon)/tensor ultralight cases are not that clear yet in the literature. 

Most of the pulsars in the PTA are located around the distances of order kpc from the Earth within a couple of wavelengths.  Fuzzy dark matter induces an oscillating pressure with the angular frequency  $\omega_c  =2 {m}$. The frequency %is \begin{align}\label{fphi}
$f_c  =\frac{\omega}{2\pi}\simeq 4.8 \text{\,nHz} \left(\frac{m}{10^{-23}\text{eV}} \right)$
%\end{align} which
 leads to the oscillation period $ T_c={f}_c^{-1} \simeq 6.6  \text{yr} \left(\frac{10^{-23}\text{eV}}{m}\right)$. As the cosmological timescale is much larger than $T_c$, the average value of the pressure is zero, and the field behaves like pressureless cold dark matter. However, here for the PTA observation, the timescale is of several decades and the oscillation in the pressure needs to be considered.

It is noteworthy that the oscillation frequencies of fuzzy dark matter models fell into the sensitive region of PTA observations.  It was already proposed earlier that the ultralight boson oscillation has a similar effect as the gravitational wave in the single pulsar timing deviations \cite{Khmelnitsky:2013lxt, Porayko:2018sfa, Nomura:2019cvc}, although for a single field dark matter candidate with the fixed mass, the signal is monochromatic. String theory naturally gives rise to more than one complex scalar field from compactification. Those stabilized fields can form an extended spectrum \cite{Easther:2005zr,Cai:2009cu,Stott:2017hvl}.  For example, if we consider extra dimensions or equivalent discretized theories like clockwork \cite{Kaplan:2015fuy,Giudice:2016yja} or dynamical dark matter \cite{Dienes:2011ja,Dienes:2016vei}, a spectrum of scalars/vectors/tensors can arise and form a spread mass spectrum. Such a particle spectrum can produce an extended frequency spectrum in timing residuals. 

%This paper is organized as follows. In section \ref{mass}, we calculate how a multi-fields spectrum can produce an extended timing residual spectrum in PTA, and discuss the extended spectra of ultralight dark matter with various spins. In section \ref{fitting}, we list and discuss the result of Bayesian analysis. In section \ref{sum}, we conclude and discuss related issues.
%In appendix \ref{1pt}, we review how an ultralight field can induce the PTA signals with a frequency-dependent monochromatic spectrum.
%In the following, we take the natural units that $\hbar=1, c=1$. 

\section{Wideband mass spectrum}
\label{mass}
It is reasonable to assume there is more than one scalar/axion field, e.g., the multiscalar/axions scenario from compactification \cite{Stott:2017hvl}. 
The mass distribution of a large number of axions is described by  random matrix theories. 
For the multiaxions as inflatons \cite{Easther:2005zr, Cai:2009cu}, the mass distribution satisfies random matrix Marcenko-Pastur law \cite{MarcenkoPastur}. 
For the fuzzy dark matter, it can also be realized in string compactification \cite{Cicoli:2014sva, Stott:2017hvl}. A typical distribution is given by the Marcenko-Pastur relation, the mass scale around $10^{-23}$eV is suppressed by the large volume.

%For the multi-fields model suggested before in Nflation \cite{Easther:2005zr, Cai:2009cu}, 
%Here, however, we do not treat axion as inflaton and fix the axion masses to be around $10^{-23}$eV. 
%This is very far from the axion mass discussed in the N-flation models. N-flation
%\subsection{Mass distributions}

We briefly discuss the mass distribution from the random matrix theory here. The mass matrix can be captured by $M =R^T R$, where $R$ is the $L\times N$ rectangular matrix. Following Marcenko-Pastur law from 1967 \cite{MarcenkoPastur},  the eigenvalue probability distribution is ${\mathcal{P}}(m^2) = {\sqrt{(m^2-m_-^2)(m_+^2-m^2)}\over{2\pi\beta \delta^2 m^2}}$, 
 for $m_-^2 \le m^2 \le m_+^2$. Here $\beta =\frac{N}{L}$, 
$m_\pm  = \delta  \left(1\pm\sqrt{\beta}\right)$ are the minimum and maximum of the mass regime. The probability density is zero outside this range, and we also have $\langle m^2 \rangle = \delta^2$. In more general cases of compactification, one can identify $\beta$ with the number of (real) axions involved in inflation divided by the total number
of moduli chiral multiplets. The value $\beta$ is rather model dependent.  For example, one can refer to \cite{Easther:2005zr,Cai:2009cu} for a discussion of the N-flation KKLT(Kachru-Kallosh-Linde-Trivedi) model \cite{Kachru:2003aw}. For the multicomponent fuzzy dark matter, we assume a similar distribution with some generic values of the parameters.

%\cite{Kachru:2003aw} 

Now we consider a spectrum of ultralight particles described by the multi-fields model with total number $N$, satisfying a distribution ${\P}(m)=dn/dm$, and $ \int  {\d}m {\P}(m) =1$. The axion number in the mass range $[m, m+dm]$ is  ${N} {\P}(m) dm$. Thus, we have $ {\P}(m) {\d}m ={\mathcal{P}}(m^2) {\d} m^2$, which leads to ${\P}(m)  = 2 m {\mathcal{P}}(m^2)$.  The  Marcenko-Pastur distribution is
\begin{align}\label{MP1}
    {{\P}}_{\beta}(m) =\frac{m}{\pi 
    \beta \delta^2}\sqrt{\left(1-\frac{m_-^2}{m^2}\right)\left(\frac{m_+^2}{m^2}-1\right)}.
\end{align}
Besides, as a comparison in the Bayesian analysis,  we also consider  the Rayleigh distribution
\begin{align}\label{MP2} 
{P}_\sigma(m)  &= \frac{m}{ \sigma^2} e^{- \frac{m^2}{2\sigma^2}}.
\end{align}
It is a distribution for nonnegative-valued random variables. In the next section, we will use these distributions for the Bayesian analysis with the observed data.

\subsection{Fuzzy dark matter}
The ultralight scalar dark matter can also be described by the summation of multifields  ${\Phi}_\phi (\bx,t) = \Sigma_I {\pa}_{I}(\bx) \cos \left[ {m_I}t + {\theta_I} (\bx) \right]$. Their oscillations will induce the perturbations in the metric. We consider the distribution of the multifields with spectral density ${\P}(m)$.
In continuous limit, the total profile for the fields is ${\Phi}_\phi (\bx,t) =\int{\d} { m} {\P}(m)\pb(m) \cos \left[ {m} t +\theta_m(\bx) \right]$, where $\theta_m$ is the random phase. It can be considered as the superposition of  dark matter waves, which are similar to the stochastic gravitational waves background from the superposition of massive black hole binary systems. The distributions of energy density and pressure are given by $\tilde \rho({m}) \simeq \frac{1}{2}m^2 {\pb(m)^2} {\P}(m)$ and $\tilde p({m}) \simeq - \tilde \rho({m}) \cos\left[2\left({m}t+ \theta_m\right)\right]$. The total energy density can be written as
$\rho_\phi\equiv \int {\d} m \tilde \rho(m) = \int {\d} m\frac{1}{2} m^2 {\pb(m)^2} {\P}(m)$.
To extend the spectrum, we can either assume ${\pb(m)^2}$ or ${\P}(m)$ as different distributions. Notice that different forms of ${\pb(m)^2}$ also depend on the thermal history of the Universe, and axion potentials. It is rather model dependent, see, e.g.,~\cite{Arias:2012az}. Here we take a convenient choice  
${\pb(m)^2}=\frac{2 \rho_\phi}{m^2}$, which leads to 
$ \tilde \rho({m}) \simeq \rho_\phi {\P}(m)$, and $\int {\d}m {\P}(m) =1$. 
We can make the ansatz for the oscillation part of the potential in the frequency space $\Psi_{osc}(\bx, t) =\int{\d} m \tilde \Psi(m) \cos\left( 2 {m}t+ 2 \theta_m\right)$. After considering the Einstein equations, we can reach the differential gravitational potential induced by the differential axions energy density $d\rho(m)$ in the mass range $[m, m+dm]$. We can use the result in \cite{Khmelnitsky:2013lxt} directly, and for each mode we have the result from the monochromatic case  $\tilde \Psi(m)   =  \frac{1}{8\mpl^2} \frac{  \tilde  \rho({m})}{{m}^2}=\frac{1}{8\mpl^2} \frac{ \rho_\phi {\P}(m)}{{m}^2}$. Considering  $m=\pi f$ and $\rho_\phi\equiv{\alpha_0} \rho_{DM}$, we reach the wideband distribution of the effective characteristic strain 
\begin{align}\label{spin0}
    h_c^{\phi}(f)  &=  \frac{{\alpha_0} }{\mpl^2} \frac{\sqrt{3}\rho_{DM}}{4\pi f}  {\P}(\pi f).
\end{align}

Other interesting cases include the spin-1 and spin-2 ultralight dark matter, which can also induce signals in the pulsar timing residuals, see, e.g., \cite{Nomura:2019cvc, Armaleo:2020yml}. One thing to notice is that both spin-1 and -2 ultralight fields are polarized to certain directions and they are known to produce anisotropic signals in pulsar timing residuals. However, for the extended frequency spectrum with multifields, we can assume their polarized directions are randomly selected and the average effect is isotropic.  Here we generalize the extended spectrum case to the spin-1 and spin-2 fields in this and the following subsection. 

\subsection{Fuzzy dark photon}
The dark photon background can produce a directional anisotropic residual signal \cite{Nomura:2019cvc} or be isotropic with a distinct signature in the two pulsar correlation. We can simply extend the spectrum of the trace part of dark photon induced residuals from previous section. The extended spectrum of the characteristic strain is then given by
  \begin{align}
    h_{c}^{A}(f) 
=  \frac{\alpha_1}{\mpl^2} \frac{\sqrt{3}\rho_{DM}}{12\pi f} {\P}(\pi f),
  \end{align}
where $\rho_A\equiv{\alpha_1} \rho_{DM}$ has been used. It is simply $ {1}/{3}$ of the scalar case $h_c^{\phi}(f)$, we will not fit them with the data later.
As we discussed already, to produce the isotropic signal, we can look at a model where the polarizations of the vector fields are randomly selected across the space. Then we need to integrate over all polarization angles. To distinguish this isotropic signal from the stochastic gravitational-wave background, we can further look at two pulsar signal correlations, known as the Hellings-Downs curve for the stochastic gravitational wave \cite{Hellings:1983fr}. One can see more discussions of these different modes and correlation curves in \cite{Armaleo:2020yml,Nomura:2019cvc}.
%Chen:2021wdo, NANOGrav:2021ini,

\subsection{Fuzzy dark graviton}
For the spin-2 ultralight dark matter model construction, one can refer to \cite{Armaleo:2020yml, Marzola:2017lbt} for the bimetric gravity and the appendix. The local oscillation of such fields with a wideband can then be $\int {\d} { m}   \mathcal{M}(m)  \cos{\left[ m t+\theta_m(\vx)\right]}\vep_{ij}\P(m)$, where $\theta_m(\vx)$ is a random phase, and $\vep_{ij}(\vx)$ is unit norm, traceless, and symmetric polarization tensor to be averaged over.  We also have the total dark matter density as $\rho_{M}= \int {\d} m\frac{1}{2}  m^2 { \mathcal{M}^2} {\P}(m)$. Take the convenient choice ${\mathcal{M}^2}=\frac{2 \rho_{M}}{m^2}$ and the replacement of $m \rightarrow 2\pi f$, we  have
    \begin{align}\label{spin2}
        h^{M}_c (f) &=\frac{\alpha_2}{\mpl} \frac{  m\mathcal{M} {\P}(m) }{\sqrt{5} }      
        =\frac{\alpha_2}{\mpl}\frac{2 \sqrt{ {\rho_{M}}}}{\sqrt{5}} {\P}(2\pi f).
    \end{align}
Notice here is different from the scalar and vector case where $m \rightarrow \pi f$, and $\rho_M $ will set to be $\rhoDM$.

\begin{figure}[h]
\centering
    \includegraphics[scale=0.5]{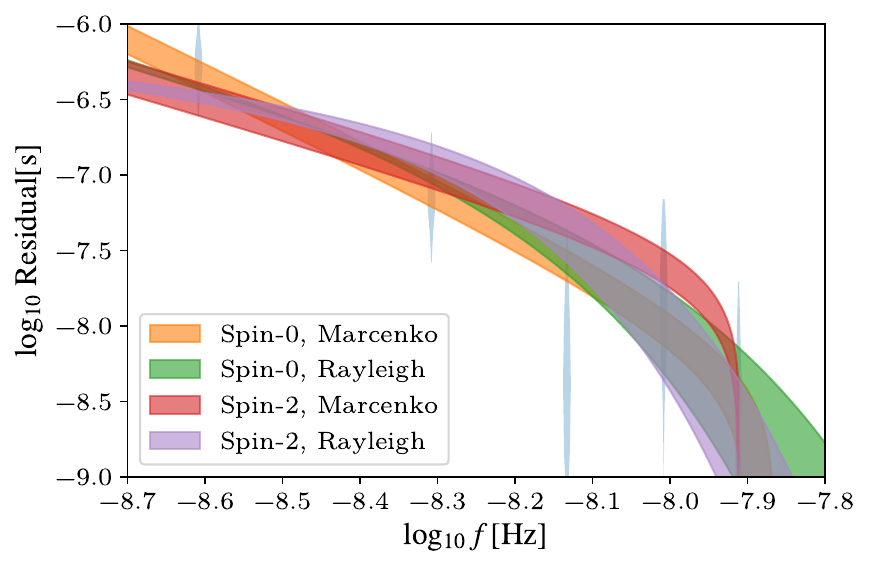}%{best_fit_inset1}
    \caption{The timing residuals due to wideband ultralight dark matter in frequency space with different distributions. With the using of Eq.~\eqref{scpf}, where $h_c (f)$ is given by Eq.~\eqref{spin0} for the spin-0 case, and  Eq.~\eqref{spin2} for the spin-2 case. The light blue violin points with error bars are reproduced from 12.5 years of NANOGrav data in \cite{NANOGrav:2020bcs}. The shaded regions indicate one sigma credible intervals. More details of the best fit lines can be found in Fig. \ref{figa1} in the Appendix.
    } \label{fig1}
\end{figure}

\section{Bayesian analysis}
\label{fitting}
\begin{figure}[h]
\centering
    \includegraphics[scale=0.48]{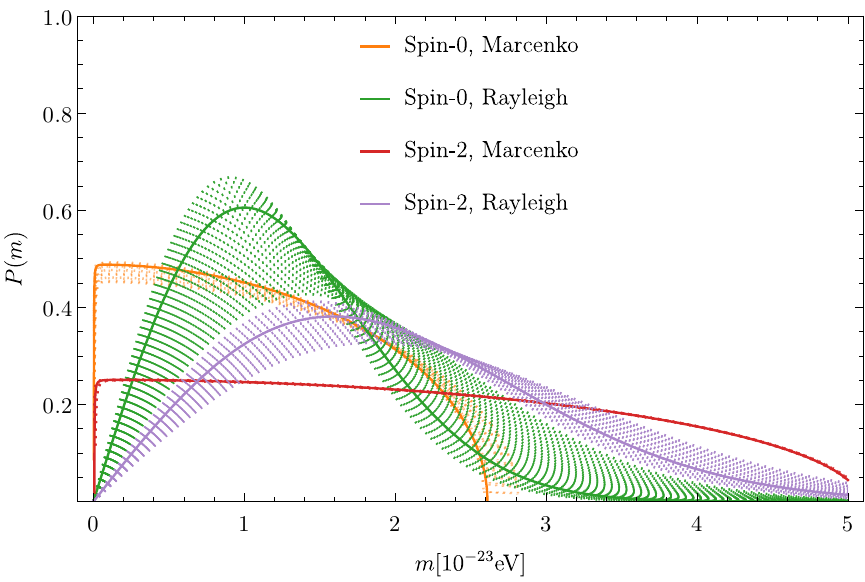}%{PDF1}
\caption{The probability distribution functions with the best fitting parameters in Table \ref{table:1}, as well as the one sigma credible intervals, which are shown by the shaded areas.
} \label{fig2}
\end{figure}

With the Marcenko-Pastur distribution in \eqref{MP1} and Rayleigh distribution in \eqref{MP2}, we perform the Bayesian analysis on the first five frequency bins of NANOGrav 12.5-yr dataset \cite{NANOGrav:2020bcs}.
%More relevant studies can be found in the references \cite{Ratzinger:2020koh}-\cite{Cai:2021dgx}.
 In Fig. \ref{fig1}, we plot the time residuals due to ultralight dark matter with spin-0 and spin-2 in frequency space with different distributions. In Fig. \ref{fig2}, we plot the probability distribution functions with different fitting parameters and distribution functions.
The best fitting results are listed in Table \ref{table:1}. 
The one sigma credible intervals are also shown in the figures and table.
CMB data and Galaxy surveys  have shown the ultralight axions below $10^{-25}$ eV can only compose a small fraction of dark matter \cite{Hlozek:2014lca,Hlozek:2016lzm,Hlozek:2017zzf,Poulin:2018dzj,Lague:2021frh,Farren:2021jcd,Dentler:2021zij,Antypas:2022asj}. 
 For the parameter choice, we consider the mass range for ultralight particles around $10^{-23}$ eV and sample $\alpha_0, \alpha_2$ as free parameters.  
\begin{table}[h!]\small\scriptsize
\centering \renewcommand{\arraystretch}{1.7}
\begin{tabular}{ |c|c|c|c|c|c|c|} 
 \hline
&Parameters~ & spin-0   &spin-1  & spin-2   \\  \hline
\multirow{3}{*}{\rotatebox[origin=c]{90}{\tiny Marcenko}}& $\alpha_i$ & $5.9 ^{+1.9}_{-1.3}$ & $\sim3\alpha_0 $ &$7.6^{+2.2}_{-1.7}\times 10^{-7}$     \\  
 \cline{2-5} % \hline
&$m_-^i$/($10^{-23}$eV) & $2.9^{+3.6}_{-0.3}\times 10^{-3}$ & $\sim\delta_0(1-\sqrt{\beta_0})$ & $6.3^{+6.0}_{-1.7}\times 10^{-3}$   \\  
 \cline{2-5} %\hline
&$m_+^i$/($10^{-23}$eV)& $2.61^{+0.21}_{-0.01} $& $\sim\delta_0(1+\sqrt{\beta_0})$ &$5.08^{+0.02}_{-0.01}$     \\  
 \hline  \hline
 \multirow{2}{*}{\rotatebox[origin=c]{90}{\tiny Rayleigh}} & $ \alpha_i$ & $5.6^{+3.8}_{-1.0}$ & $\sim3\alpha_0$ & $6.1^{+2.1}_{-1.3} \times 10^{-7}$ \\  \cline{2-5}% \hline
& $\sigma_i$/($10^{-23}$eV) & $1.0^{+0.4}_{-0.1} $ & $\sim\sigma_0$ &$1.6^{+0.3}_{-0.1} $   \\    \hline
\end{tabular}
\caption{The best fitting model parameters with one sigma credible intervals. For the Marcenko-Pastur distribution in Eq.\eqref{MP1}, the minimum and maximum of the mass regime $m_\pm^{i}  = \delta_i  \left(1\pm\sqrt{\beta_i}\right)$ are the derived parameters, the sampled parameters are $\alpha_i$,  $\beta_i$, and $\delta_i$.
For the Rayleigh distribution in Eq.\eqref{MP2}, the sampled parameters are 
$\alpha_i$ and $\sigma_i$.
}
\label{table:1}
\end{table}

The results in Table~\ref{table:1} show that if we use the wideband spin-0 model to fit the NANOGrav data, around $6\rho_{DM}$ is required. In other words, we can put the constraints of the wideband spin-0 model at this level, which is similar to \cite{Porayko:2018sfa} for the single mass case. 
Since we consider a wide spectrum of particles, the bounds from Lyman-alpha can be relaxed and it would be interesting to perform a detailed analysis with the observation in the future.
%so we don’t have this problem.
%we do not require $\alpha_0 = \rho_\phi/\rho_{DM} $ to be strictly one in the fitting. Also notice that the best fitted spectrum of spin-1 particle is identical to the spin-0 case, with 3$\alpha_0$ as the overall factor.
In fact, the NANOGrav signals can mainly come from the gravitational waves quadrupolar contribution. In this sense, a small monopole component from scalar dark matter is still possible. 
%The spin-0 model can still survive with a small contribution to the timing residuals. 
%In this sense, we cannot rule out the scalar case. 
For the spin-2 case, due to the tensor structure and multiple modes in \eqref{spin2}, the constraints and parameters are different. We can set the whole density as $\rho_{DM}$, and the coupling $\alpha_2$ is adjustable as in~\cite{Armaleo:2020yml}.
It would be interesting to perform more detailed data analysis and observational constraints on the spin-2 ultralight dark matter models in future work.

 % and conclusion
\section{Discussion}
\label{sum}
The ultralight dark matter can produce the pulsar timing signals at the monochromatic frequency. Here we generalize it to an extended spectrum, provided a multifields scenario. Then the induced effects are similar to the isotropic stochastic gravitational-wave background in PTA. We also discuss the higher spin cases with vector/dark photon oscillation and spin-2 oscillation. We can see different signals: either with an angular dependence in single pulsar timing residuals as suggested in \cite{Nomura:2019cvc}, or isotropic signal with a special curve in two-point timing residuals correlation, provided that the background oscillation polarizations are randomly selected over the galactic scale \cite{Chen:2021wdo, NANOGrav:2021ini}. The later isotropic cases for both spin-1 and spin-2 can mimic the stochastic gravitational-wave background. The ultralight dark matter can also arise from emergent and holographic scenarios \cite{Cai:2017asf, Cai:2018ebs, Bigazzi:2020avc, Huang:2021aac,Li:2021qer}, which would be interesting to study their signals in PTA. For other modes and correlation curves in the metric and pulsar timing array, see, e.g., \cite{DeMartino:2017qsa,Kato:2019bqz,Cyncynates:2021yjw,Chen:2021bdr,Demirtas:2021gsq,Xue:2021xts}.

In summary, we discuss pulsar timing residuals from the ultralight dark matter oscillation. Especially, we generalize the study with a single frequency to an extended spectrum.  Then the ultralight dark matter can also be a viable candidate explaining the claimed common-spectrum  at nano-Hz in the NanoGrav 12.5 years dataset.  This study provides more interpretations for the NANOGrav results and more means to identify the ultralight dark matter in current and future PTA data. In the future,  FAST \cite{Nan:2011um} and SKA \cite{Kramer:2015jsa} can further probe or constrain such wideband ultralight dark matter.
 
\section*{ Acknowledgments}%\small
We thank many valuable suggestions from Professor~R.-G. Cai, and helpful discussions with Y. Gao, Z.-K. Guo, J. Liu, S. Pi. This work was supported by the National Natural Science Foundation of China (Grants Nos. 12005255, 12105013, 11690022, 11821505, 11851302, 11947302, 11991052), the National Key Research and Development Program of China (Grant No.  2021YFC2201901), the Key Research Program of the Chinese Academy of Sciences (CAS Grant No.~XDPB15), the Key Research Program of Frontier Sciences of CAS.
%, Xiao Xue and Xing-Jiang Zhu

\appendix
\section{Oscillation induced timing residuals}
\label{1pt}
We review the induced timing residual from single field ultralight dark matter. 
%The ultralight particle with a mass around $10^{-23}$ eV oscillating in space acts as an interesting dark matter candidate, the fuzzy dark matter.
The ultralight dark matter can be described by the action 
$S = \Sigma_{(s)}\int {\d}x^4 \sqrt{-g}\, {\mathcal L}_{(s)}$, where $s=0,1,2$ represent the scalar, vector and tensor fields, respectively. And the energy momentum tensor is given by $T_{\mu\nu} = \frac{-2}{\sqrt{-g}}\frac{\delta S}{\delta g^{\mu\nu}}$.
The ultralight particle oscillation induces the oscillation of the energy-momentum tensor, so as to trigger the oscillation in the metric. As an approximation, we consider the flat background, with the perturbed metric in the Newtonian gauge
\begin{equation}\label{metric0}
  ds^2 =- \left(1 + 2 \Phi\right) {\d}t^2 + \left[\left(1 - 2 \Psi \right) \delta_{ij}+h_{ij} \right] {\d}x^i {\d}x^j.
\end{equation}
From the Einstein equations $G_{\mu\nu}=\frac{1}{\mpl^2} T_{\mu\nu}$, the trace parts of the perturbations lead to
\begin{align}
-2 \nabla ^2 \Psi &= \frac{1}{\mpl^2}{T^t}_{t},\label{hEQ1} \\
2 \ddot{\Psi}  +\frac{2}{3} \nabla^2 (\Phi-\Psi) &=\frac{1}{3\mpl^2} {T^k}_{k},\label{hEQ2}
\end{align}
where the spatial derivatives $\nabla^2\equiv \delta^{ij}\partial_i \partial_j$.
The traceless part of the metric perturbation $h_{ij}$ satisfies
\begin{align}
\ddot{h}_{ij}-\nabla^2 {h}_{ij}&= \frac{2}{\mpl^2}\left(T_{ij}- \frac{1}{3} \delta_{ij} {T^k}_k\right).\label{hEQ3}
\end{align}
Here we focus on dark matter oscillations across the coherent length at the kpc scale and neglect the cosmic expansion.

\subsection{Spin-0: Massive scalar field}
We begin with the massive scalar field $\phi$, with the Lagrangian density
\begin{align}\label{Lspin0}
{\mathcal L}_{\phi} =-\frac{1}{2}(\partial \phi)^2 - \frac{1}{2}{m^2} \phi^2.
\end{align} 
Assuming the profile of the scalar field
$\phi (\bx,t) =  {\pa}(\bx) \cos \left[ {m}t + {\theta_0} (\bx) \right]$, with the random phase $\theta_0(\bx)$ and plugging this ansatz into the stress energy tensor. After dropping the subdominant part $(\nabla \phi )^2 \sim  {m^2}v^2 \phi^2$, we obtain the energy density $ \rho_{\phi}  \equiv -{T^t}_{t} \simeq \frac{1}{2}m^2 {\pa}^2$, and pressure $p_{\phi}   \equiv \frac{1}{3}{T^k}_{k} \simeq - \rho_{0} \cos\left[2\left({m}t+  \theta_\phi(\bx)\right)\right]$. This oscillation in the pressure induces the oscillation in the gravitational potentials. In the metric \eqref{metric0}, it is enough to consider the leading oscillating contributions from cosine type  $\Psi \simeq \bar{\Psi}(\bx) + \Psi_\phi \cos\left[2\left({m}t+   {\theta_0}(\bx)\right)\right]$, and similarly for $\Phi$ in the metric \eqref{metric0}. After considering the linearized Einstein equations and $\rho_\phi={\alpha_0} \rho_{DM}$, we can reach
\begin{align}\label{psi1}
    \Psi_\phi&=  \frac{1}{8\mpl^2} \frac{ \rho_{\phi} }{{m^2}}
\simeq    6.5\times 10^{-16} {{\alpha_0}}
 \left(\frac{10^{-23}\text{eV}}{m}\right)^{2}.
\end{align}

For an pulse sent at the point $\{{\x}_{0},{t_{0}}\}$ with frequency ${\omega_{0}}$ and detected at the Earth $\{{\x}_\phi, t_\phi\}$ with with frequency $\omega_\phi(t)$, the Doppler effect leads to
\begin{align}\label{eqSW}
 z_\phi(t)&\equiv \frac{{\omega_{0}}-\omega_\phi(t)}{{\omega_{0}}} \simeq \Psi(\x_\phi,t_\phi) - \Psi({\x}_{0},{{t_{0}}}).
\end{align}
Considering the small velocity $v \simeq k/\omega  \sim 10^{-3}$,  we have neglected higher order terms as discussed in \cite{Khmelnitsky:2013lxt}. 
For the scalar oscillation, the change of frequency can induce the timing residual in the pulse
$R_\phi(t) =  \int_0^t z_\phi(t') d t'  = \tilde{r}_{\phi}
    \cos{\left( 2{m}t+   \theta_{\phi} + {\theta_0} -{m} D\right)}$, with the amplitude
$\tilde{r}_{\phi}  \equiv  \frac{\Psi_\phi}{m}\sin{\left({m} D + \theta_{\phi} - {\theta_0} \right) }$ and $D\equiv |\x_\phi-\x_0|$. The average  square value of  pulsar-timing residual is
\begin{align}\label{rab0}
    \Delta t_\phi &= \sqrt{ \langle \tilde{r}_\phi \tilde{r}_\phi  \rangle} 
    \simeq \frac{1}{\sqrt{2}} \frac{  \Psi_\phi}{m} %\nn\\&
\simeq 30\text{ns} \left(\frac{10^{-23}\text{eV}}{m}\right)^{3} \alpha_0.
\end{align}
It is induced by the oscillation of gravitational potential background $ \Psi_\phi$ in \eqref{psi1}.

We can compare this with the timing residuals caused by a monochromatic gravitational wave signal with the strain $h_\phi$  and amplitude $ \tilde{r}_h = \frac{h_\phi}{\omega} \sin{\l \frac{\omega D (1 - \cos{\theta})}2 \r} (1 + \cos{\theta}) \sin{(2\psi)}$
\cite{Jenet:2003ew,Wen:2011xc}.
After integrating over $\theta, \phi, D$, we have
\begin{align}\label{Rab}
    \Delta t_h= \sqrt{\langle \tilde{r}_h\tilde{r}_h   \rangle} \simeq \frac{1}{\sqrt{6}} \frac{ h_\phi}{\omega}.
\end{align}
The factor of $\frac{1}{2} $ from Hellings-Downs relation has been considered.
We can identify \eqref{rab0} with \eqref{Rab}. Considering $\omega=2m$, we obtain
\begin{align}\label{hcphi}
    h_\phi & = 2 \sqrt{3} \,\Psi_\phi    
    =\frac{\sqrt{3}}{4\mpl^2} \frac{ \rho_{\phi} }{{m^2}} % \nn\\&
 \simeq 5.2 \times10^{-17} \alpha_0
    \left(\frac{\fy}{f}\right)^{2}  ,
\end{align}
where the frequency  $f= m/\pi$ and  $\rho_\phi={\alpha_0} \rho_{DM}$ have been considered.
Thus, the scalar field dark matter can mimic the monochromatic gravitational wave in PTA.
For the spin-1 and spin-2 ultralight dark matter, we briefly summarize the results as below.

\subsection{Spin-1: Massive vector field}
For the massive vector field $A_\mu$, we consider the Lagrangian density
\begin{align}\label{Lspin1}
{\cal L}_{A} = -\frac{1}{4} F^2- \frac{1}{4} m^2 A^2.
\end{align}
The oscillating solution of the vector field is given by
\begin{align}
A_\mu= \mathcal{A}_\mu \cos{[m  t+\theta_1(x)]},
\end{align}
where $\theta_1(\vx)$ is a random phase.
Such background oscillation can also induce oscillation in the trace part as well as a traceless part in the metric \eqref{metric0}.

For ultralight vector particle, the trace part contributes to the timing residuals, which do not depend on the polarization of the ultralight vector field. Its magnitude is $1/3$ of the scalar case, as derived in \cite{Nomura:2019cvc},
\begin{align}
{h}_A % = \frac{1}{3}{h}_\phi 
= \frac{\sqrt{3}}{12\mpl^2}\frac{ \rho_A}{ m^2}\, .
\end{align}
Thus, the vector field dark matter can also produce the pulsar timing signal at the monochromatic frequency.

For the traceless part of spacial perturbation $h_{ij}$ in  the metric \eqref{metric0}, following the calculation in \cite{Nomura:2019cvc}, the photon propagating null geodesic with the perturbed metric is
$  \frac{d \xi^\mu }{d \lambda}=-\Gamma^\mu_{ab} \xi^a\xi^b$.
Then we have the modified frequency at the linear level:
$\frac{d \xi^0 }{d \lambda}=-\frac{1}{2} \dot{h}_{ij} \omega_0^2 n^i n^j$, where $\omega_0$ is the unmodified frequency and $n^i$ is the unit three-vector. 
After the integration, we can arrive at
\begin{equation}\label{AA}
z_A (t )\equiv \frac{\omega_0- w_A(t)}{\omega_0} =\frac{1}{2}n^i n^j [h_{ij} (t_A)-h_{ij}(t_0)]. 
\end{equation}
For the traceless part, the residual depends on the polarization direction as
$ h_{A}^{\theta} = \frac{\sqrt{3}}{6 \mpl^2} (1+3 \cos 2\theta)\frac{\rho_A }{ m^2}$ \cite{Nomura:2019cvc}. 

The vector oscillation as discussed in \cite{Nomura:2019cvc} has a preferred direction at each spatial point. It is interesting to discuss whether all the polarization directions aligned or not within a coherent patch. Notice that the current pulsar timing array roughly spans the same length as the ultralight dark matter coherent length at the kpc scale. If vector polarization has a preferred direction, then we can see it from the single pulsar timing residues, as pointed out in \cite{Nomura:2019cvc}. If the polarization directions for the vector oscillation are stochastic and randomly selected at each point, it is useful to discuss the average isotropic effect. We can then average over the sphere $(\theta, \phi )$ of all possible directions, and  the traceless part above averages to zero.

\subsection{Spin-2: Massive tensor field}
For a massive spin-2 field $\M_{\mu\nu}$, according to the dark matter candidate in bimetric gravity \cite{Armaleo:2020yml}, we consider the Fierz-Pauli Lagrangian density
\begin{align}\label{Lspin2}
{\mathcal L}_{M} = \frac{1}{2} \M_{\mu\nu}\mathcal{E}^{\mu\nu\rh\si}\M_{\rh\si} - \frac{1}{4} m^2 \left( \M_{\mu\nu}\M^{\mu\nu} - \M^2 \right) ,
\end{align}
where $\mathcal{E}^{\mu\nu\rh\si}$ is the Lichnerowicz operator, which is defined via
$\mathcal{E}^{\mu\nu}_{~~\rho\sigma} \equiv  \delta^\mu_\rho \delta^\nu_\si \Box - g^{\mu\nu} g_{\rh\si} \Box + g^{\mu\nu} \D_\rho \D_\si 
+ g_{\rh\si} \D^\mu \D^\nu - \delta^\mu_\si \D^\nu \D_\rho - \delta^\mu_\rho \D^\nu \D_\si$, and $\M = g^{\mu\nu} \M_{\mu\nu}$.
The oscillating solution of the spin-2 field is given by
\begin{align}
M_{ij}&\, =\mathcal{M}  
    \cos{[m  t+\theta_2(x)]}\vep_{ij},
\end{align}
where $\theta_2(\vx)$ is a random phase, and $\vep_{ij}$ is a symmetric and traceless polarization matrix with unit norm. 

Now if we consider its impact on the pulsar arriving residual, one can refer to the interaction as below.
\begin{align}
    S_\text{int} = -\frac{\alpha_2}{2\mpl} \int d^4x\,\sqrt{-\tilde{g}} M_{\mu\nu} T^{\mu\nu} \,,
\end{align}
where ${\alpha_2}$ characterizes the mixing parameter between the spin-2 field and the matter field.
The contribution of  $M_{0\mu}$  are in higher order and higher derivative, so we neglect them and work with
$\tilde{g}_{ij} =  \delta_{ij} + \frac{{\alpha_2}}{\mpl}{M}_{ij}$.
For a free photon with unperturbed four-momentum $\xi^\mu=\omega_0(1, n^i)$, 
we have the geodesics
$\frac{\dd \xi^0}{\dd \lambda}  %= -\Gamma^0_{ij} \xi^i \xi^j
	 = -\frac{{\alpha_2}\omega_0^2}{2\mpl } \dot{M}_{ij} n^i n^j$,
where $\lambda$ is the affine parameter.  Then keeping only the linear terms in $\alpha_2$ and performing the integral, we have
\begin{align}\label{eqfreq0}
z_M (t)\equiv \frac{\omega_0- w_M(t)}{\omega_0}
=  \frac{{\alpha_2}}{2\mpl }n^i n^j [M_{ij} (t_M)-M_{ij}(t_0)],
% =  \frac{{\alpha_2}}{2\mpl } \int_0^t d t \omega_0 \partial_t{M}_{ij}  n^i n^j,  
\end{align} 
where \(\omega_0\) is the unperturbed frequency at the pulsar. 

One can follow the treatment in bimetric gravity, absorb $M_{\mu\nu}$ in the metric and calculate the delay in the pulses arriving time.  For the stochastic background we also need to average over celestial sphere of the spin-2 polarization. The average  square value of  pulsar-timing residual is $  \Delta t_M %=\sqrt{ \langle R^2_2(t)\rangle } 
\simeq \frac{{\alpha_2} \mathcal{M}}{\sqrt{30}\mpl } \sin\left(mt+\theta_2\right)$  \cite{Armaleo:2020yml}.
This corresponds to the gravitational strain 
\begin{align}
{h}_M = \frac{{\alpha_2} }{\mpl}\frac{\mathcal{M} }{\sqrt{5}} 
= \frac{{\alpha_2} }{\mpl} \frac{\sqrt{2\rho_M}}{\sqrt{5} m }.
\end{align}
Here $\rho_M $ can be set as before $\rhoDM=0.4$~GeV/cm$^3$.

More relevant studies on ultralight dark matter can be found in the references \cite{Ratzinger:2020koh}-\cite{Ferko:2021bym}.
In Figs. \ref{figb2} and \ref{figb3}, we show the results of Bayesian analysis on the first five frequency bins of NANOGrav 12.5-yr dataset for the spin-0 and spin-2 ultralight dark matter with Marcenko-Pastur and distribution, respectively.

\begin{widetext}

\begin{figure}[h!]
 %   \centering
    \includegraphics[scale=0.5]{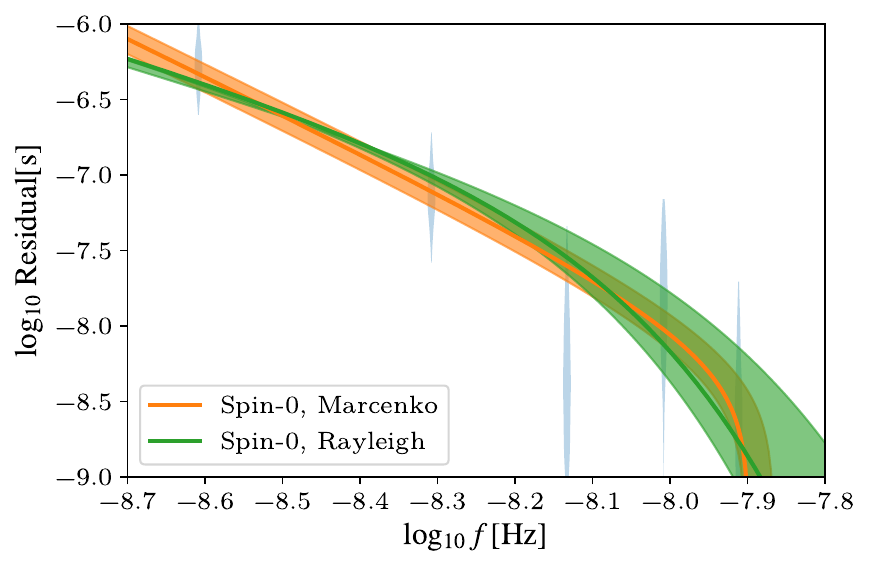}\qquad\qquad  
    \includegraphics[scale=0.5]{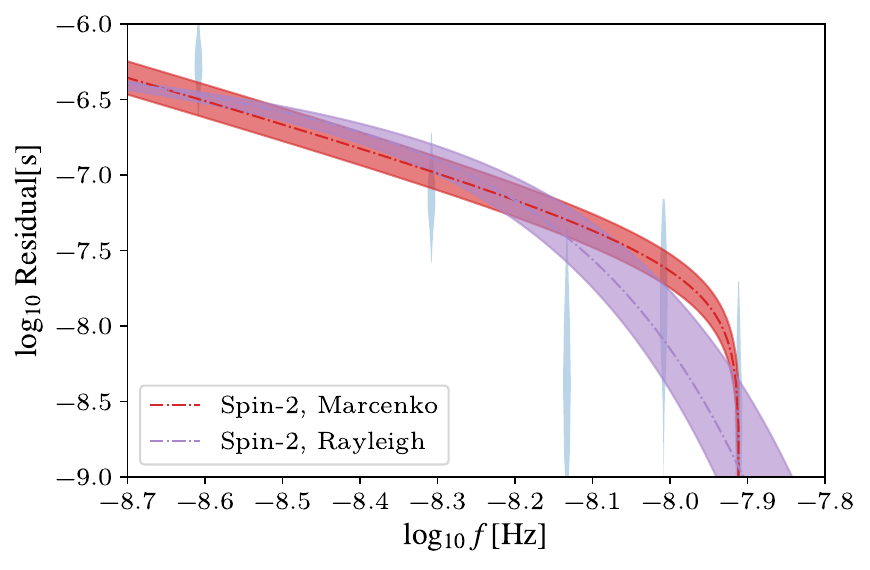}
    \caption{The timing residuals due to in frequency space with the first five frequency bins of NANOGrav 12.5-yr dataset. Left: the best fit line with one sigma credible intervals of the spin-0 ultralight dark matter with both distributions.  Right: the best fit line with one sigma credible intervals of the spin-2 ultralight dark matter with both distributions.} \label{figa1}
%    Here ${\alpha_0}\equiv\rho_\phi/\rho_{DM}$
\end{figure}

\begin{figure}[h!]
 % \centering
    \includegraphics[scale=0.45]{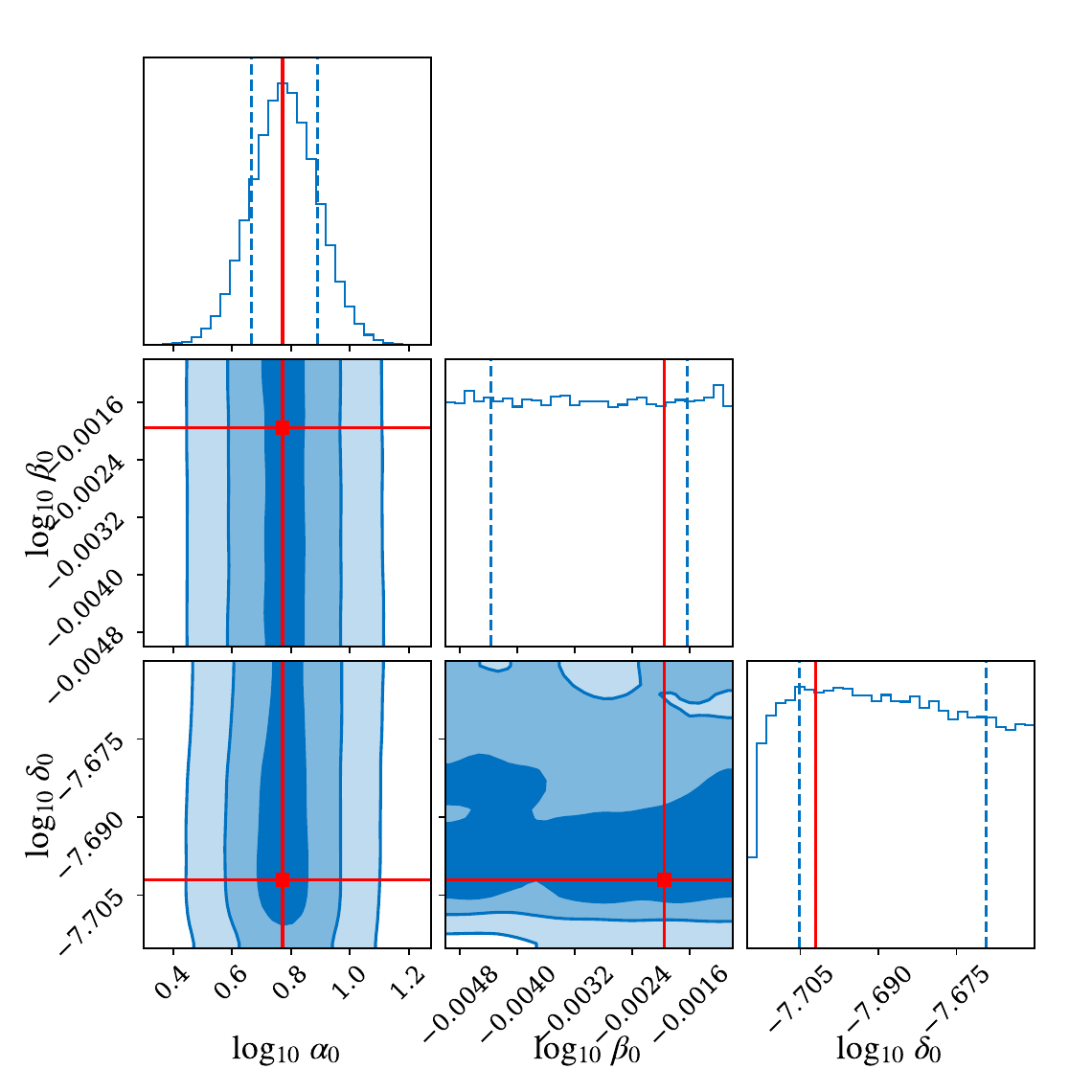}
    \includegraphics[scale=0.45]{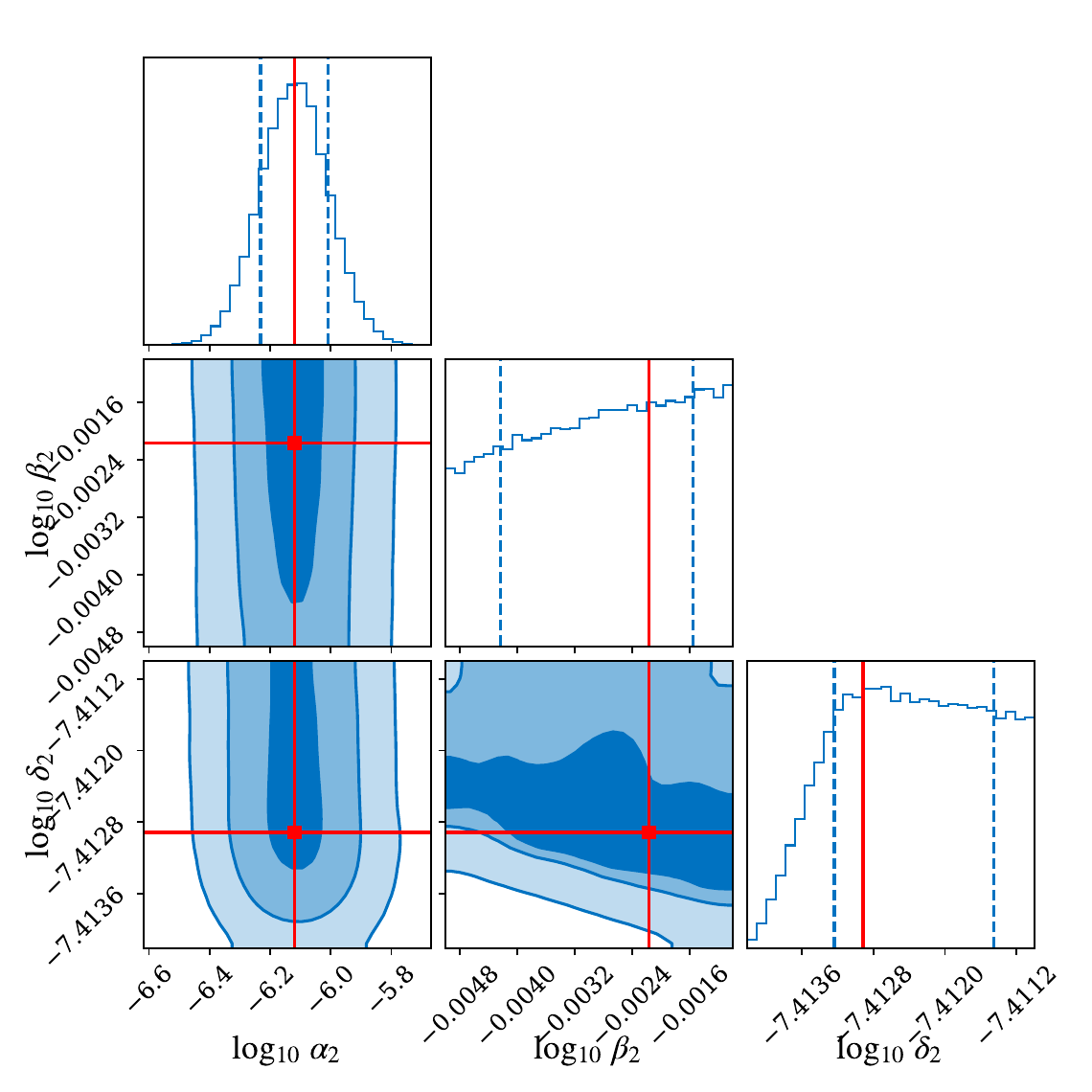}
    \caption{The Bayesian analysis results on the first five frequency bins of NANOGrav 12.5-yr data for the spin-0 and spin-2 ultralight dark matter with Marcenko-Pastur distribution. % \eqref{MP1}.
Here $\alpha_i$ and $\beta_i$ are dimensionless, $\delta_i$ is plotted with the unit of Hz.   
}\label{figb2}
%Here ${\alpha_2}$ is the interaction parameter between dark graviton and matter.
\end{figure}

\begin{figure}[h!]
 %   \centering
    \includegraphics[scale=0.5]{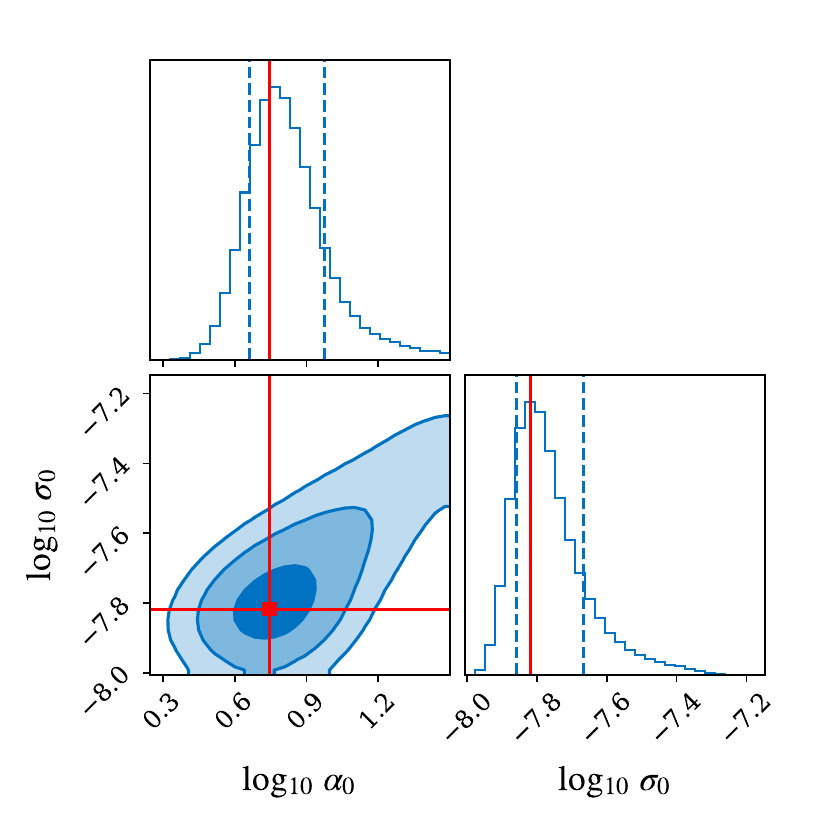}\qquad\qquad  
    \includegraphics[scale=0.5]{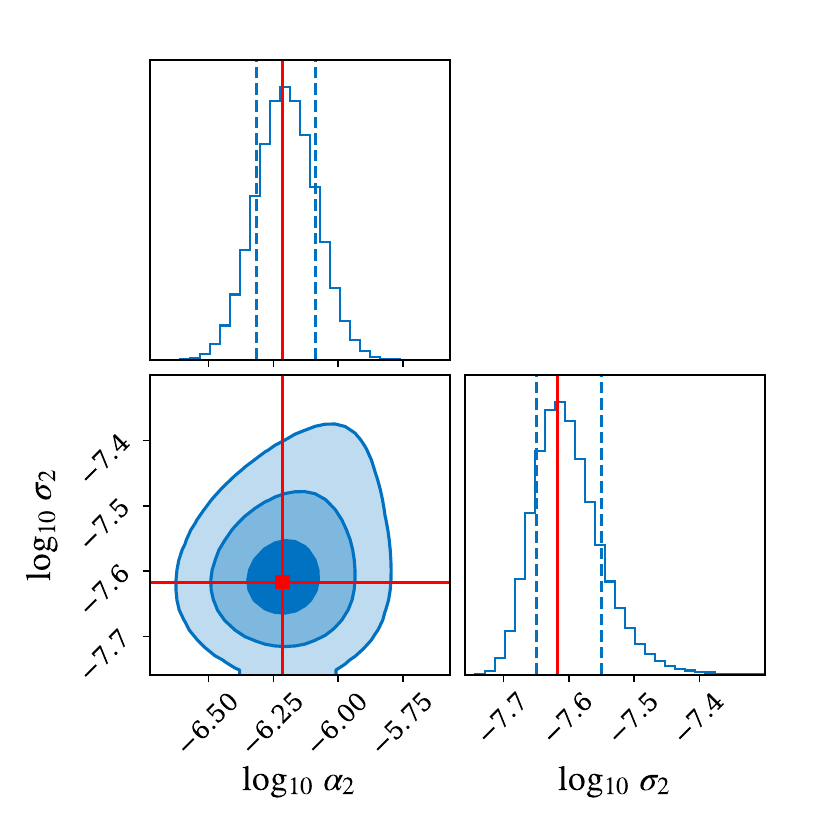}
    \caption{The Bayesian analysis results on the first five frequency bins of NANOGrav 12.5-yr data for the spin-0 and spin-2 ultralight dark matter with Rayleigh  distribution. % \eqref{MP2}. 
Here $\alpha_i$ is dimensionless, $\sigma_i$ is plotted with the unit of Hz.
} \label{figb3}
%    Here ${\alpha_0}\equiv\rho_\phi/\rho_{DM}$
\end{figure}

\end{widetext}

\end{document}